\documentclass{ecai} 

\usepackage{latexsym}
\usepackage{amssymb}
\usepackage{amsmath}
\usepackage{amsthm}
\usepackage{booktabs}
\usepackage{enumitem}
\usepackage{graphicx}
\usepackage{color}

\newcommand{\BibTeX}{B\kern-.05em{\sc i\kern-.025em b}\kern-.08em\TeX}

\begin{document}


\begin{frontmatter}


\paperid{123} 


\title{D-CNN and VQ-VAE Autoencoders for Compression and Denoising of Industrial X-ray Computed Tomography Images}


\author[A]{\fnms{Bardia}~\snm{Hejazi}\orcid{0000-0002-9262-128X}\thanks{Corresponding Author. Email: bardia.hejazi@bam.de.}}
\author[A]{\fnms{Keerthana}~\snm{Chand}\orcid{0009-0000-5748-2264}}
\author[A]{\fnms{Tobias}~\snm{Fritsch}\orcid{0000-0002-0472-2874}}
\author[A,B]{\fnms{Giovanni}~\snm{Bruno}\orcid{0000-0001-9632-3960}} 

\address[A]{Department for X-Ray Imaging, Federal Institute for Materials Research and Testing (BAM), Berlin, Germany}
\address[B]{Institute of Physics and Astronomy, University of Potsdam, Potsdam, Germany}



\begin{abstract}

The ever-growing volume of data in imaging sciences stemming from the advancements in imaging technologies, necessitates efficient and reliable storage solutions for such large datasets. This study investigates the compression of industrial X-ray computed tomography (XCT) data using deep learning autoencoders and examines how these compression algorithms affect the quality of the recovered data. 
Two network architectures with different compression rates were used, a deep convolution neural network (D-CNN) and a vector quantized variational autoencoder (VQ-VAE). The XCT data used was from a sandstone sample with a complex internal pore network. The quality of the decoded images obtained from the two different deep learning architectures with different compression rates were quantified and compared to the original input data. In addition, to improve image decoding quality metrics, we introduced a metric sensitive to edge preservation, which is crucial for three-dimensional data analysis. We showed that different architectures and compression rates are required depending on the specific characteristics needed to be preserved for later analysis. The findings presented here can aid scientists to determine the requirements and strategies for their data storage and analysis needs.

\end{abstract}

\end{frontmatter}


\section{Introduction}

With the advancement of imaging technologies, the quality and size of data have increased significantly, leading to a corresponding rise in storage requirements. The expanding volume of data and the complexity of its processing present ongoing challenges for industry~\citep{Brady2019, Sagiroglu2013} and science alike~\citep{Succi2019}. Scientific research, in particular, contributes substantially to the overall volume of big data. For instance, the Large Hadron Collider (LHC) at CERN generates an immense amount of data. The LHC raw data estimates, without any trigger selection, can reach 40,000 exabytes per year, which is approximately 80 times the total data stored by Amazon Web Services~\citep{Clissa2023}. Similarly, other research areas, such as medical and biological sciences, produce large quantities of high-resolution images, further contributing to the data deluge. Managing the ever-growing influx of information is a significant challenge that highlights the need for advanced data storage and processing technologies to ensure that valuable information is accurately preserved and accessible for analysis.

X-ray computed tomography (XCT) is a significant source of large data in both medical and industrial settings. Traditional compression methods have been used to compress imaging data in the form of lossless and lossy compression~\citep{Vijayvargiya2013}. Lossless compression retains all the information from the original data, ensuring that the recovered data is identical to the original. Common lossless compression formats, such as PNG and GIF, achieve varying degrees of compression based on the structure of the original data. Alternatively, lossy compression sacrifices some information, resulting in a recovered image that is similar, but not identical, to the original. Traditional compression techniques have been applied to compress three-dimensional (3D) XCT images, with wavelet compression being used for medical datasets~\citep{Schelkens2003}. Wavelet compression has also been utilized in industrial XCT applications, including data denoising~\citep{Stock2020, Lang2023}. Additionally, JPEG compression has been explored for industrial XCT, examining its impact on the metrological quality of recovered data~\citep{Kieß2025}.

Recently, deep learning autoencoders have emerged as a powerful compression method that often outperform traditional techniques in terms of compression rates and recovered image quality~\citep{Cheng2018, Petscharnig2017}. Autoencoders are a type of lossy compression that consist of two sections, an encoder and a decoder network. The encoder network, typically a convolutional neural network (CNN), maps the input data to a lower-dimensional latent space. The decoder network, essentially the inverse of the encoder, reconstructs the compressed data back to its original form.

Autoencoders are a form of unsupervised learning that can learn the underlying structure of data without needing any explicit labels. Learning the underlying structure of the data allows for more efficient compression and reconstruction. Additionally, this capability enables autoencoder models to denoise input data. Commonly used autoencoders are based on CNN architectures and consist of convolution layers that encode the image to a lower dimensional latent space, and deconvolution layers that decode the compressed data to restore the image. Previous studies have demonstrated the potential of CNN autoencoders in compressing and denoising medical X-ray images~\citep{Florian2025, Kwon2020, Senapati2022, Sushmit2019, Fettah2024}.

Other autoencoder architectures, such as Variational Autoencoders (VAE), have also been used to compress medical images~\citep{Liu2022}. VAEs encode a continuous, probabilistic representation of the latent space, typically sampled from a Gaussian distribution. Consequently, in addition to compression and denoising, VAEs can utilize the variational inference to generate new data similar to the sampled original input data~\citep{Kingma2019}.

Vector-Quantized Variational Autoencoders (VQ-VAEs) are another type of autoencoder that operates on a discrete latent space~\citep{Oord2018}. Using a discrete latent space simplifies the optimization problem as compared to learning a continuous distribution in the case of VAEs, which is usually more challenging. In a VQ-VAE, a discrete codebook is initialized and developed by sampling categorical distributions of the posterior and prior. This codebook is then used to decode the encoded data. Such a discrete representation is beneficial for generative models, where real-world data is often discrete in nature, such as images of different objects that do not reside in a continuous space.

Applying deep learning methods to the compression of industrial XCT data presents unique challenges. Unlike the medical field, where it may be straightforward to establish a general database for human X-ray images, because of the abundance and similarity of the images available, industrial applications and materials are too diverse for a generalized compression model. Furthermore, the quality of the recovered data is crucial, as a detailed analysis of the properties of industrial XCT samples requires each pixel in the recovered image to be as similar to the original input data as possible. Consequently, previous studies have not extensively explored the compression of industrial XCT data using deep learning methods.

In this study, we addressed these challenges by applying deep learning techniques to compress industrial XCT data obtained from a sandstone sample. We demonstrated that this approach could effectively reduce data size while maintaining the quality of the recovered output for subsequent analysis. We investigated the performance of two different network architectures, a deep convolution neural network (D-CNN) similar to ones used in previous studies investigating the compression of medical X-ray images, and a state-of-the-art vector quantized variational autoencoder (VQ-VAE). We also considered the performance of the models at different compression rates to determine the impact of the model architecture and the compression rate on the output data quality. We quantified the performance by calculating parameters that may be of interest to scientists, such as the total porosity percentage. Additionally, we introduced an image quality metric that was sensitive to edge preservation, which is an important feature to preserve for the analysis of 3D volumetric datasets. The work presented here can provide valuable insights for scientists engaged in applied and industrial applications, offering guidance on optimal compression schemes that balance the need for reduced data storage with the preservation of data quality.

\section{Methods}

\begin{figure}[tbh]
\centering
\includegraphics[width=8cm]{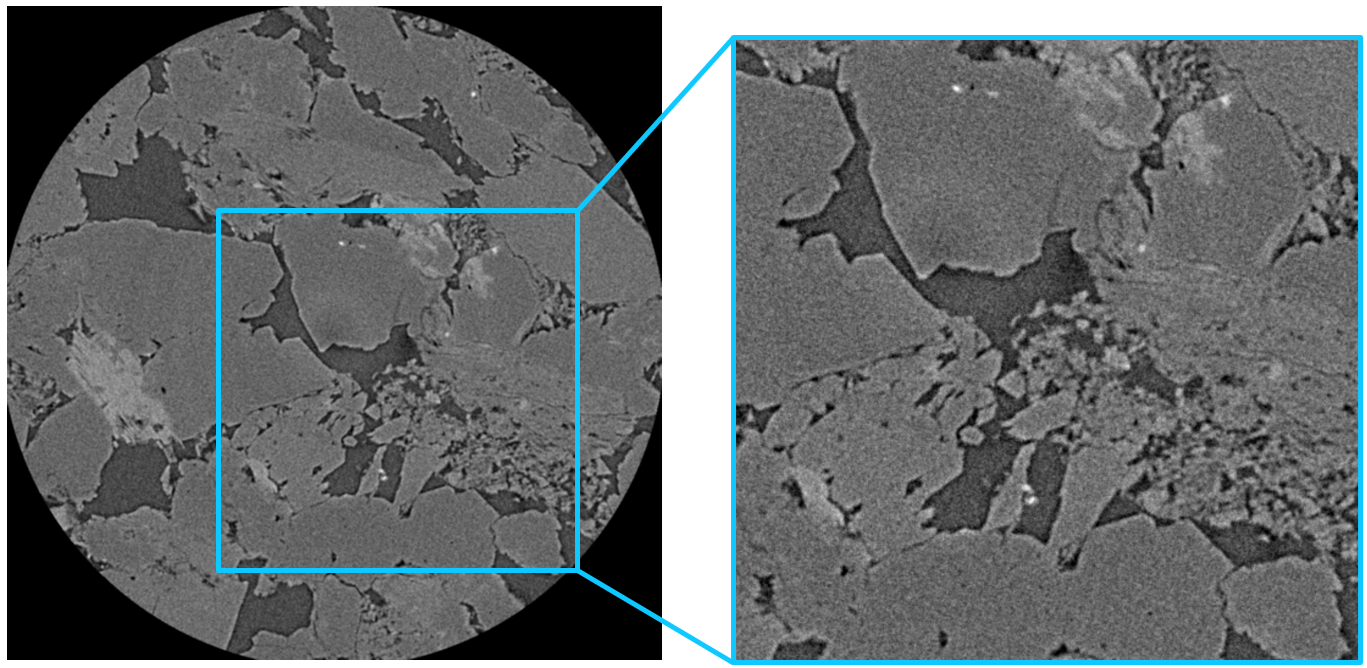}
\caption{Example slice from the XCT data of sandstone with original dimensions of $998 \times 998$ pixels and the cropped section used for the compression analysis with $512 \times 512$ pixels.}
\label{fig:sample_example}
\end{figure}

The XCT images used in this work were of a Kylltaler sandstone sample with a diameter of 37 mm that was scanned using the ZEISS Xradia 620 Versa XCT machine. The X-ray tube operated at a voltage of 80 kV with a power of 10 W. The total number of projections was 6400 with a 5 s detector exposure time per projection. The sample was imaged using the $20\times$ magnification objective and placed at a focus-object-distance (FOD) of 12.0 mm. The focus-detector-distance (FDD) was 23.0 mm. The images used were from slices along the vertical axis of the volumetric data and had an original size of $998 \times 998 \times 969$ with a voxel size of 0.7 $\mu$m. The original images were cropped to a size of $512 \times 512 \times 900$, as shown in Fig.~\ref{fig:sample_example}. Using smaller sized images allowed us to achieve faster computation speeds and were also well suited for structuring deep learning model layers. Additionally, with the cropped area we only considered the interesting features that are important in the analysis of such materials, such as the pore space, and did not include the background surrounding the sample in the compression models. Fig.~\ref{fig:sample_example} shows one slice of the reconstructed XCT data. The XCT images (i.e. reconstruction slices) of the sandstone sample were chosen as a representative test case to consider for image encoding and decoding of materials with complex and detailed inner structures.

\begin{figure}[tbh]
\centering
\includegraphics[width=8.5cm]{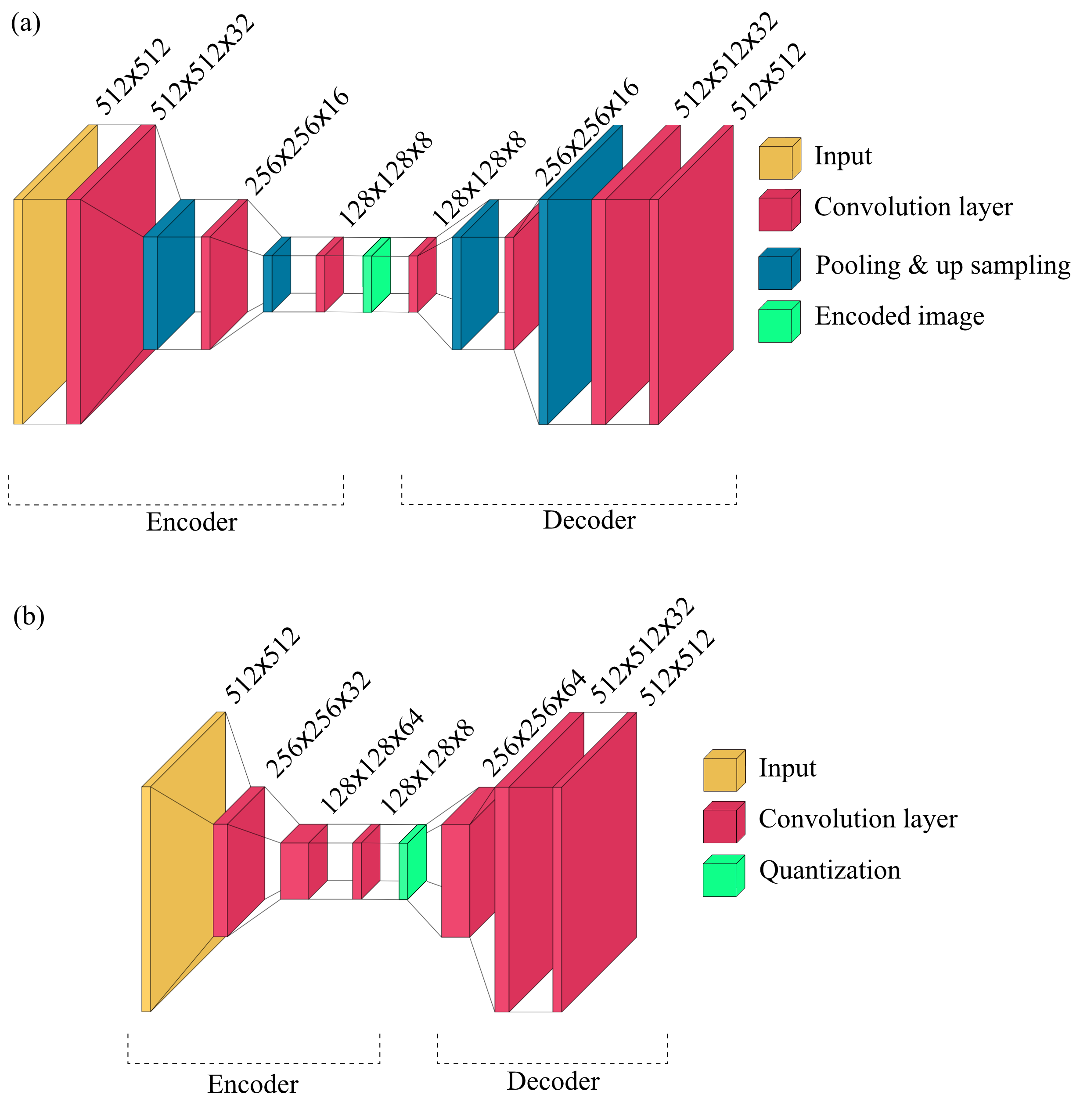}
\caption{Schematic of the architectures used to compress $512 \times 512$ input images to an $128 \times 128 \times 8$ encoded array. (a) CNN autoencoder showing the encoder and decoder with details of their convolution, pooling and up sampling layers. (b) VQ-VAE autoencoder showing the encoder, decoder, and quantization layer (Visualization done using visualkeras package~\citep{Gavrikov2020}).}
\label{fig:models}
\end{figure}

We used two architectures for compression, a deep convolutional neural network (D-CNN) model and a vector quantized variable autoencoder (VQ-VAE) model. The two types of architecture used are shown in Fig.~\ref{fig:models} with an example of the model layers for a compression from initial images with size $512 \times 512$ to an encoded form with size $128 \times 128 \times 8$. The D-CNN model used is shown in Fig.~\ref{fig:models} (a). The convolution layers in the D-CNN model had ReLU activation functions with kernel size 3. The convolution layers were followed by Max pooling layers in the encoder section and up-sampling layers in the decoder section. The final output convolution layer in the decoder had a Sigmoid activation function. The D-CNN model was trained using the Adam optimizer and binary cross-entropy loss function.

\begin{table}[tbh]
\caption{Number of total trainable parameters in the models examined in this study.}
\centering

\begin{tabular}{c|c@{\hspace{8mm}}ll} 

Model & Total parameters \\
 \toprule
 
$128\times128\times8$ CNN & 12785\\
$64\times64\times4$ CNN & 12937\\
$32\times32\times2$ CNN & 12977\\
$128\times128\times8$ VQ-VAE & 43273\\
$64\times64\times4$ VQ-VAE & 190981\\
$32\times32\times2$ VQ-VAE & 781315\\
\bottomrule

\end{tabular}
\label{table:model-params}
\end{table}

The VQ-VAE model was based on the original formulation by Oord~et.~al.~\citep{Oord2018,Paul2021} and the model architecture is shown in Fig.~\ref{fig:models} (b). The VQ-VAE consisted of an encoder, quantization layer, and decoder. Both the convolution layers of the encoder and the deconvolution layers of the decoder had a kernel size of 3 and stride size of 2. The convolution layers had ReLU activation functions except for the last convolution layer in both the encoder and decoder sections, which had no activation functions. The loss function used was the mean squared error between the original image and the decoded image.

In addition to the models shown in Fig.~\ref{fig:models} with a compression to a $128 \times 128 \times 8$ encoded array, we also investigated compressions to $64 \times 64 \times 4$ and $32 \times 32 \times 2$ encoded arrays. In the case of the larger compressions, the model architectures were similar to those shown in Fig.~\ref{fig:models} where for the compression to $64 \times 64 \times 4$ we added one and for the compression to $32 \times 32 \times 2$ we added two additional convolution and deconvolution layers.
The input data was normalized and divided into an 80/20 training and validation dataset with fixed random state for reproducibility. The training was then performed for 100 epochs with a batch size of 128 for all the models and compressions considered. The number of trainable parameters in each model is shown in Table~\ref{table:model-params}. 

After decoding the encoded images, we quantitatively investigated the quality of the recovered images. We calculated different image quality metrics to determine how well the fine features and overall characteristics of the material are preserved. 
The calculations were done on the training dataset since the main goal of this work was to evaluate the performance of the models in learning the input data and how accurately the input data is reproduced from its compressed form.

\section{Results}

\begin{figure*}[tbh]
\centering
\includegraphics[width=17cm]{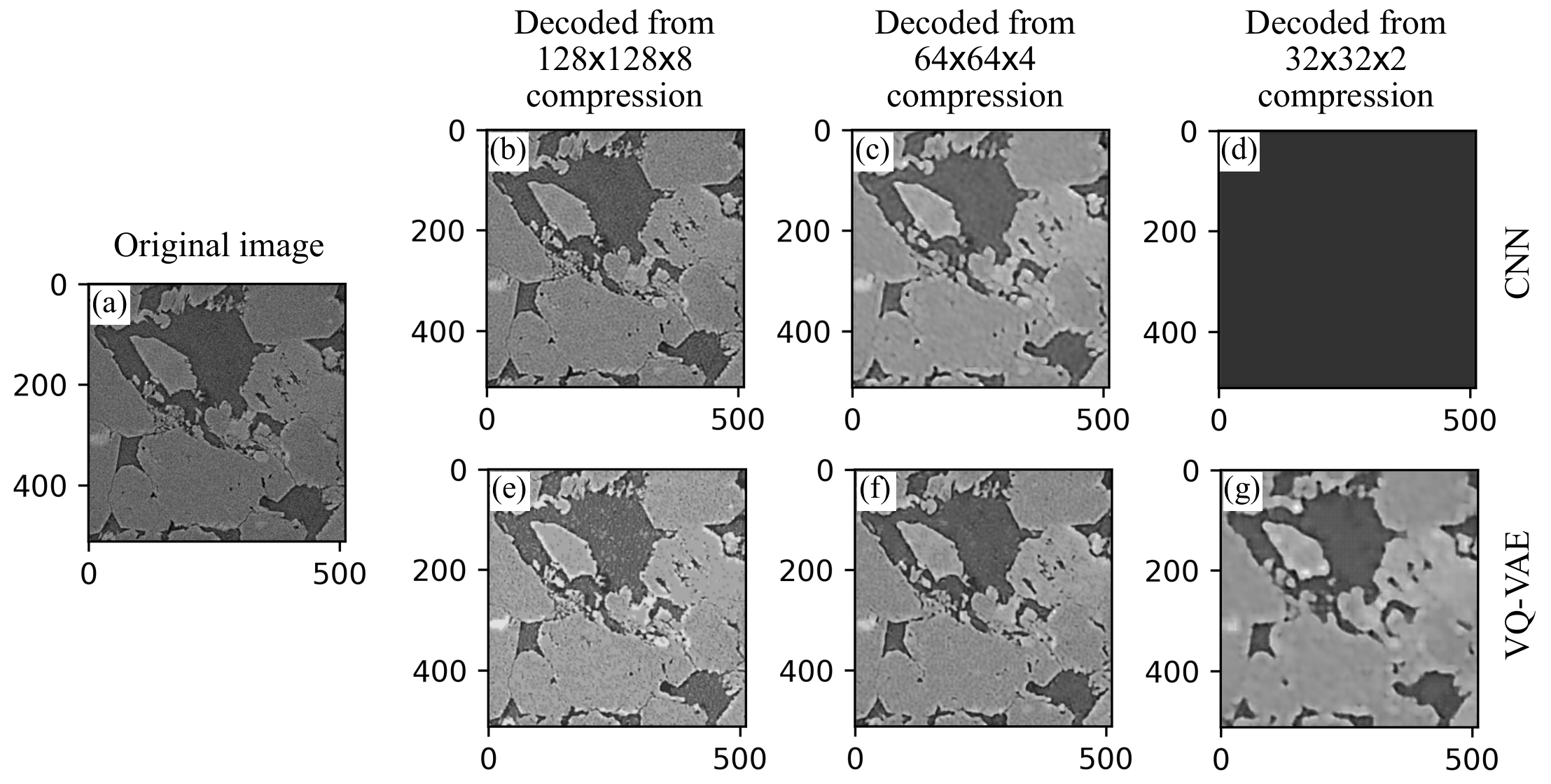}
\caption{(a) Original $512 \times 512$ 8-bit image. Output images obtained from the (b,c,d) CNN and (e,f,g) VQ-VAE autoencoder with different compression rates. The compressions to $128 \times 128 \times 8$, $64 \times 64 \times 4$, and $32 \times 32 \times 2$ encoded arrays were examined. The quality of the output images decrease as the compression rate increases for the two architectures.}
\label{fig:8bit-compression}
\end{figure*}

The output decoded images for the two different model architectures and three compression rates along with the original image slice from XCT are shown in Fig.~\ref{fig:8bit-compression}. The original image is an 8-bit grayscale image slice consisting of mainly two distinct grayscale values. The two main phases present in the material are the actual sandstone material (light gray areas) and pore network (dark gray areas). Other materials with different absorption factors (gray values) may also be present but in lesser quantities.

\begin{table*}[tbh]
\caption{The mean square error (MSE) and peak signal to noise ratio (PSNR) of the 8-bit images. The quality metrics are shown for the D-CNN and VQ-VAE models at compressions to $128 \times 128 \times 8$, $64 \times 64 \times 4$, and $32 \times 32 \times 2$ encoded arrays. The compression to $32 \times 32 \times 2$ is only shown for the VQ-VAE model since the D-CNN model failed to learn appropriate decoding at this compression rate.}
\centering

\begin{tabular}{c|c|c|c|c|c @{\hspace{8mm}}ll} 
 & $128\times128\times8$ CNN & $64\times64\times4$ CNN & $128\times128\times8$ VQ-VAE & $64\times64\times4$ VQ-VAE & $32\times32\times2$ VQ-VAE \\
 \toprule
MSE & 0.03 & 0.042 & 0.037 & 0.04 & 0.048 \\
PSNR (dB) & 63.29 & 61.84 & 62.37 & 62.0 & 61.25 \\
\bottomrule
\end{tabular}

\label{table:8bit-quality}
\end{table*}

As the compression rate increases, the decoded images lose some of the finer features of the original image. In the case of the D-CNN decoded image for the largest compression (Fig.~\ref{fig:8bit-compression} (d)) to a $32 \times 32 \times 2$ encoded array, the autoencoder is unable to learn the structure and fails to decode the image entirely. The VQ-VAE autoencoder is still able to decode the image and capture the overall structure at the largest compression to a $32 \times 32 \times 2$ encoded array, although the decoded image is blurry and does not capture the fine structure (Fig.~\ref{fig:8bit-compression} (g)).

\begin{figure*}[tbh]
\centering
\includegraphics[width=14cm]{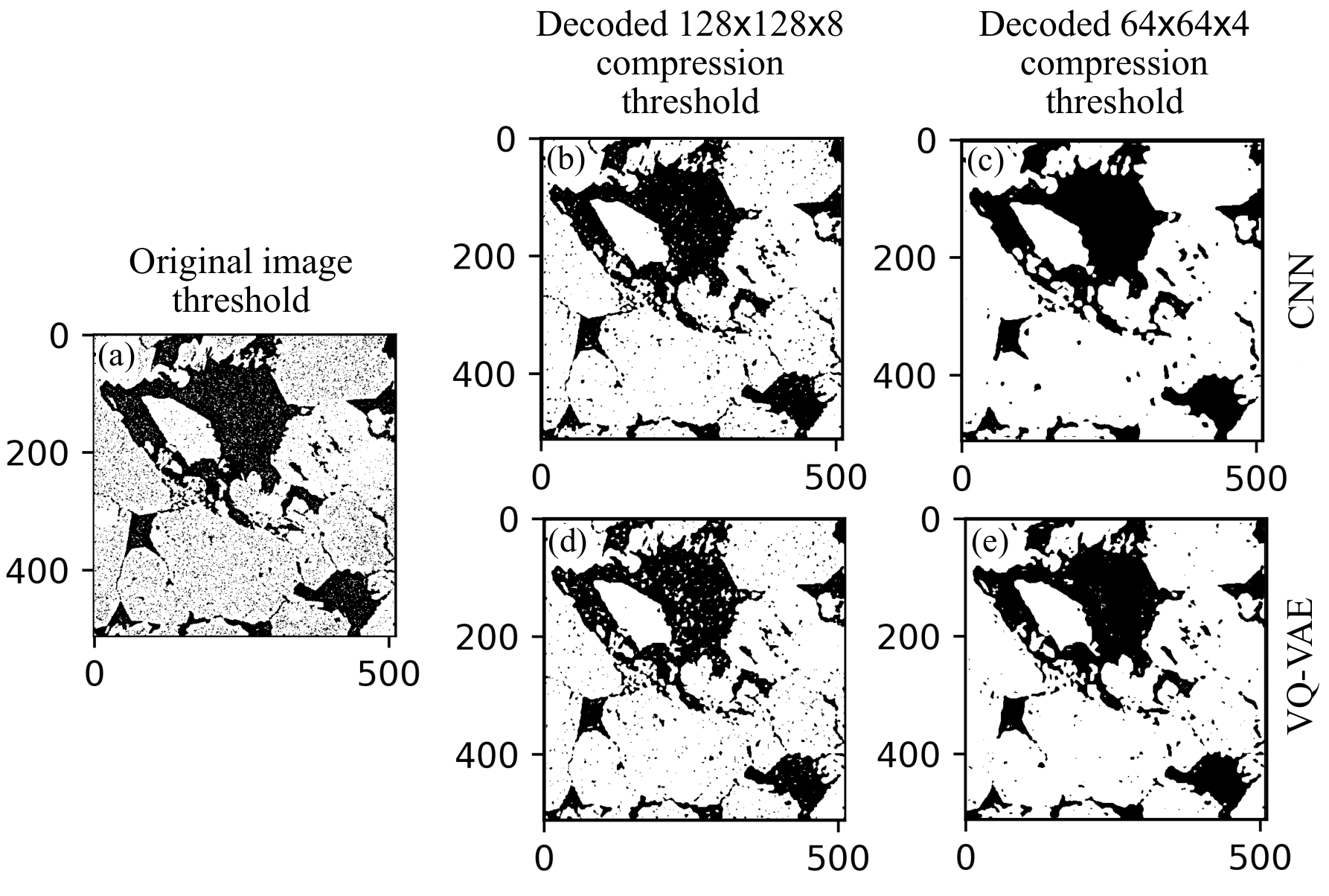}
\caption{(a) Lower Otsu threshold of the original 8-bit image. (b,c) Lower Otsu threshold of the decoded image obtained from the D-CNN model. (d,e) Lower Otsu threshold of the decoded image obtained from the VQ-VAE model. The output images had an increased denoising effect as the compression rate also increased.}
\label{fig:8bit-threshold}
\end{figure*}

To compare the two models and different compression rates quantitatively and assess the decoded image quality, we calculated the most commonly used metrics in literature, the mean square error (MSE) and peak signal to noise ratio (PSNR)~\citep{Ferreira2024}. The MSE is given by

\begin{equation}
MSE=\frac{1}{mn}\sum_{i=0}^{m-1}\sum_{j=0}^{n-1}\left[ A(i,j)-B(i,j) \right]^{2}.
\end{equation}
Here, the original image, $A$, and the decoded image, $B$, have the same size $m \times n$. The PSNR is calculated using the MSE and is defined as,

\begin{equation}
    \begin{split}
    PSNR &= 10 \log_{10}\left(\frac{MAX_{A}^{2}}{MSE}\right) 
    \\
    &=20\log_{10}\left(MAX_{A}\right) -10\log_{10}\left(MSE\right).
    \end{split}
    \label{eq:psnr}
\end{equation}
The term $MAX_{A}$ is the maximum intensity value able to be assigned to a given pixel (e.g. 255 for an 8-bit image). 

The values for the MSE and PSNR of the 8-bit decoded images obtained from the models and compression rates investigated are reported in Table~\ref{table:8bit-quality}. The quality metrics are not reported for the $32 \times 32 \times 2$ compression for the D-CNN model since the decoding failed in that case. The quality metrics do not significantly change between the different models and compression rates considered. The MSE changed by $<2\%$ at most from the worst to best performing case and similarly the PSNR only changed by approximately 2 dB. 

The D-CNN model performed only slightly better than the VQ-VAE model at the smaller compression to $128 \times 128 \times 8$.
However, at larger compression rates the VQ-VAE performed better than the D-CNN model. The better performance from the VQ-VAE model is expected, since at the larger compression rate, the VQ-VAE model had a significantly larger number of total trainable parameters and thus the model was able to better capture the underlying data structure. For the case of the $64 \times 64 \times 4$ compression, the VQ-VAE had approximately $15\times$ the total number of trainable parameters as compared to the D-CNN model. The models also behaved similarly at the largest compression to $32 \times 32 \times 2$, where the D-CNN with a low amount of total trainable parameters failed, but the VQ-VAE model was still able to recover the overall structure by having approximately $60\times$ the number of total trainable parameters.

Since the compression to $32 \times 32 \times 2$ resulted in improper decoding and significant loss of detail, we did not consider the $32 \times 32 \times 2$ compression in the remainder of this work and only considered the two smaller compression rates in further analysis and comparisons.

\begin{figure*}[tbh]
\centering
\includegraphics[width=14cm]{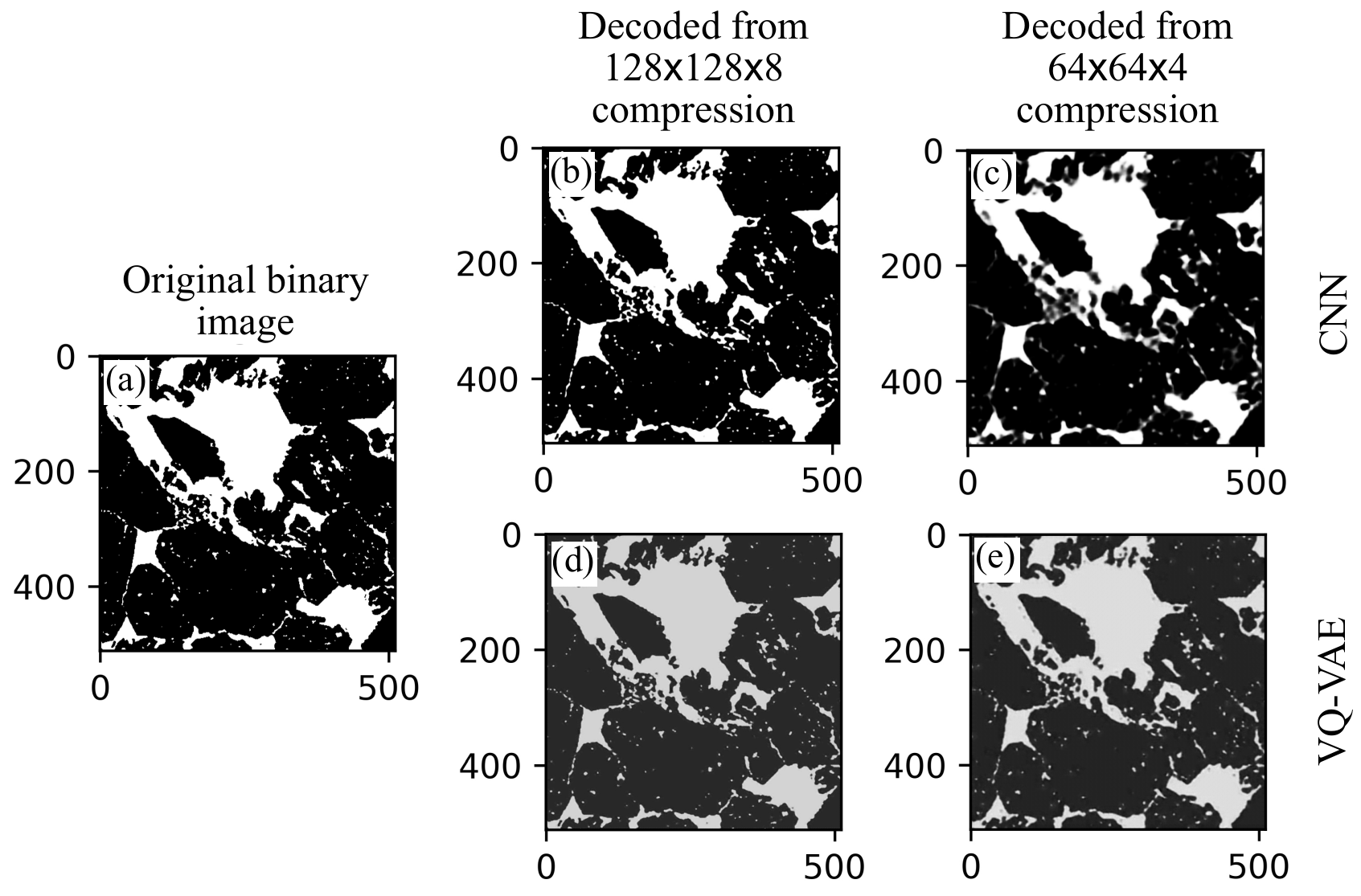}
\caption{(a) Original binarized image. Output images obtained from the (b,c) D-CNN and (d,e) VQ-VAE models with the different compression rates.}
\label{fig:binary-compression}
\end{figure*}

In addition to compression, autoencoders are used to reduce image noise by learning the underlying structure in images~\citep{Bajaj2020,Gondara2016,Vincent2008}. To assess the denoising achievable by the models and compression rates investigated in this study, we threshold the 8-bit decoded images and the original image of Fig.~\ref{fig:8bit-compression} using Otsu's method~\citep{Otsu1979}. The threshold images are shown in Fig.~\ref{fig:8bit-threshold}. As shown in Fig.~\ref{fig:8bit-threshold} (a), the sandstone material in the original image is quite noisy, while the noise decreased as the compression rate increased in the decoded images. As a result, while the increase in compression rate reduced the ability of the autoencoder to preserve fine features, an advantage of larger compression rates was that the noise in the image was reduced.

\begin{table*}[tbh]
\caption{Porosity percentage of the original binarized dataset as compared to the decoded images obtained from the two models and different compression rates investigated.}
\centering

\begin{tabular}{c|c|c|c|c|c@{\hspace{8mm}}ll} 

 & Original & $128\times128\times8$ CNN & $64\times64\times4$ CNN & $128\times128\times8$ VQ-VAE & $64\times64\times4$ VQ-VAE \\
 \toprule
Porosity $\%$ & 19.16 & 19.84 & 18.95 & 19.06 & 18.99 \\
\bottomrule
\end{tabular}
\label{table:porosity}
\end{table*}

To further quantify the quality of the decoded images and calculate parameters of importance to scientists (e.g. the porosity of porous media)~\citep{Xiong2016}, we simplified the input original image by binarizing the data into the two main phases: the sandstone material and the pore network. The binarized original image is shown in Fig.~\ref{fig:binary-compression} (a), where a mean shift filter was applied to the 8-bit image and a manual threshold was applied to reduce and minimize any noise and obtain a clean input image that clearly displays the two distinct phases. The binarized data was then used to train the two model architectures with the two different compression rates and the resulting decoded images are shown in Fig.~\ref{fig:binary-compression} (b-e).

\begin{table*}[tbh]
\caption{The MSE, PSNR, and mean square Laplacian error (MSLE) of binarized images. The quality metrics are shown for the D-CNN and VQ-VAE models at compressions to $128 \times 128 \times 8$ and $64 \times 64 \times 4$.}
\centering

\begin{tabular}{c|c|c|c|c@{\hspace{8mm}}ll} 
& $128\times128\times8$ CNN & $64\times64\times4$ CNN & $128\times128\times8$ VQ-VAE & $64\times64\times4$ VQ-VAE \\
 \toprule
MSE & 0.01 & 0.034 & 0.011 & 0.025 \\
PSNR (dB) & 19.99 & 14.64 & 19.61 & 15.98 \\
MSLE & 0.019 & 0.06 & 0.021 & 0.048 \\
\bottomrule

\end{tabular}
\label{table:binary-quality}
\end{table*}

The overall structure of the input data was well preserved in the models and compression rates examined, as seen by the decoded images. However, as the compression rate increased, the blurriness in the decoded images also increased, especially in the case of the D-CNN model architecture, as shown in Fig.~\ref{fig:binary-compression} (c). To quantify the decoding precision and better evaluate the decoding quality, we calculated the total porosity percentage of the original input data and the decoded data obtained from the models and compression rates investigated. The porosity percentage was calculated as the total number of pore network pixels (white pixels in Fig.~\ref{fig:binary-compression}) divided by the total number of pixels. The porosity percentage is shown in Table~\ref{table:porosity}. All models are performing well in capturing the overall structure and porosity present in the sample since the porosity percentage difference between decoded data and original data is $<1\%$. Such a porosity percentage difference is within the experimental error 
or it can also occur among repeated measurements of the same sample,
depending on the segmentation procedure employed~\citep{Carmignato2018}.

Interestingly, at the smaller compression to $128\times128\times8$, the total porosity obtained from the VQ-VAE model is closer to the original total porosity as compared to the total porosity obtained from the D-CNN model. However, the D-CNN model at the smaller compression to $128\times128\times8$ had a smaller $MSE$ and larger $PSNR$. From Fig.~\ref{fig:binary-compression} and Table~\ref{table:porosity} we can conclude that if the information needed for analysis and later recovery is the overall structure of the material and quantities such as total porosity of the material, then the largest compression rates can be applied to conserve the most amount of data storage space. Since the total porosity percentage difference between decoded images and original image is small for all models and compression rates considered,

\begin{figure}[tbh]
\centering
\includegraphics[width=7cm]{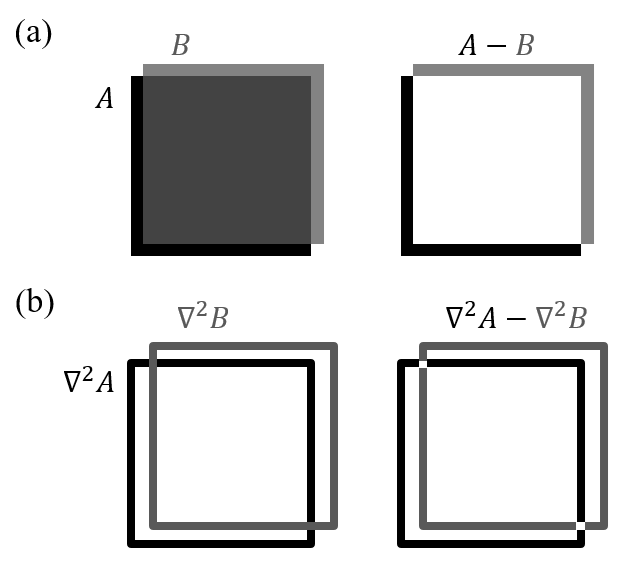}
\caption{Schematic comparing how the (a) mean square error (MSE) and the proposed (b) Laplacian difference calculate misalignment in image features. The Laplacian difference can allow us to determine how well features and edges are preserved in decoded images.}
\label{fig:laplacian-sketch}
\end{figure}

The quality metrics for the binary decoded images obtained from the different models and compression rates considered are shown in Table~\ref{table:binary-quality}. The D-CNN and VQ-VAE both perform similarly for the smaller compression to $128 \times 128 \times 8$. As expected, the VQ-VAE model outperforms the D-CNN model in the quality of the decoded images for the larger compression to $64 \times 64 \times 4$. 

Previous studies have shown that the MSE and the PSNR are not ideal metrics for image quality assessment~\citep{Ferreira2024,Quan2012,Wang2004,Zhang2018,Lin2025}. These metrics are not sensitive to blur in images as shown in our calculations, where the MSE for 8-bit images does not significantly change while the images are clearly blurred at the larger compression rates. The MSE and PSNR treat all errors equally and are not necessarily sensitive to structural features to which humans may be more sensitive, such as edges and textures. It is also worth noting that in the case of binary images $MAX_{A}=1$ and so the first term in Eq.~\ref{eq:psnr} is zero, thus the $PSNR$ is essentially providing similar information about the quality as the $MSE$.

\begin{figure*}[tbh]
\centering
\includegraphics[width=17cm]{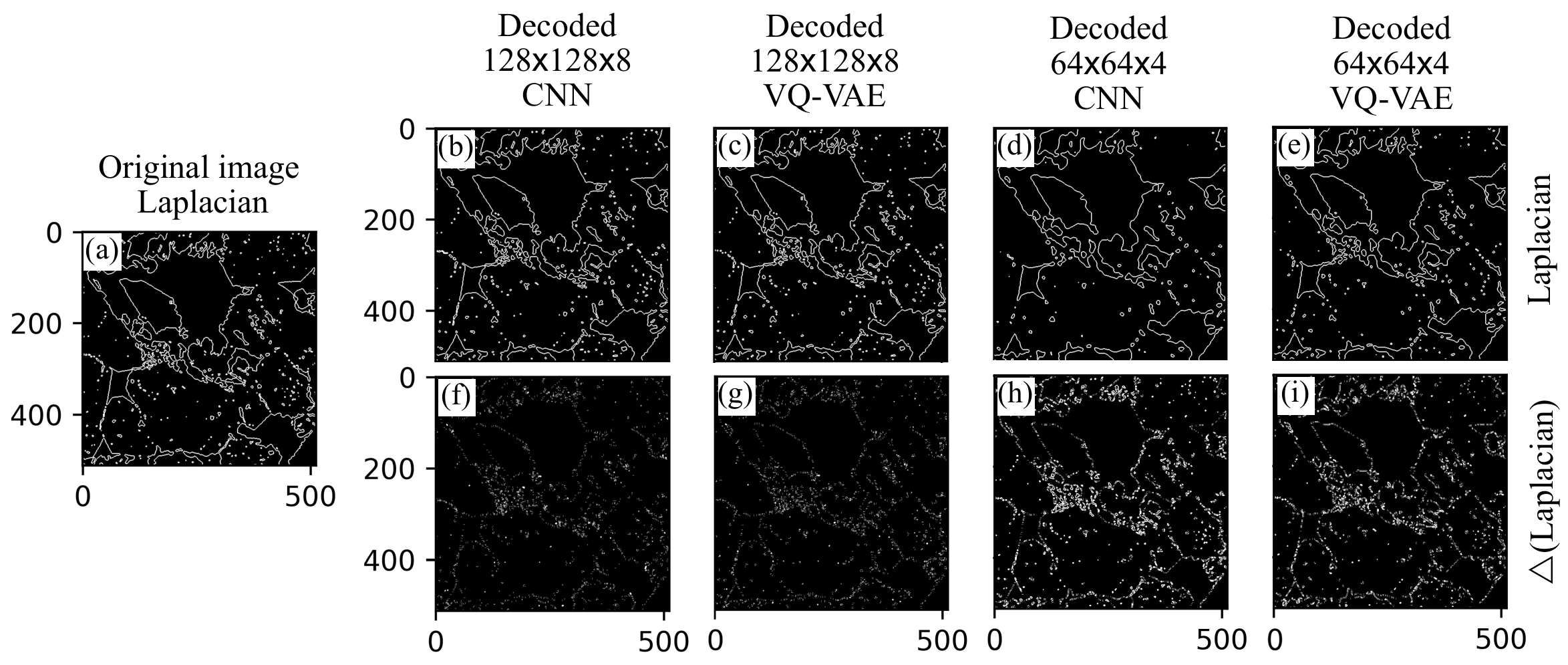}
\caption{(a) Laplacian of the original binarized images revealing the edges of the features present in the image. (b,c,d,e) Laplacian of the output images from the models and compression rates studied. (f,g,h,i) Difference image between the Laplacian of the original image and output images showing how well the features are conserved during compression.}
\label{fig:laplacian}
\end{figure*}

To better quantify the decoding quality on a finer scale, we propose the use of a new metric that is sensitive to edges and image fine structures, the mean square Laplacian error (MSLE). The MSLE is defined as 

\begin{equation}
MSLE=\frac{1}{mn}\sum_{i=0}^{m-1}\sum_{j=0}^{n-1}\left[\nabla^{2} \left( A(i,j)-B(i,j) \right) \right] ^{2},
\end{equation}
where $\nabla^{2} = \frac{\partial^{2}}{\partial x^{2}} + \frac{\partial^{2}}{\partial y^{2}}$ is the Laplacian along the $x$ and $y$ directions of an image. The MSLE is sensitive to edge alignment between images, so misalignment in structures such as translations or scalings and smoothing of edges can be better detected. Such changes to feature edges are of importance in 3D data analysis since they can possibly change any metrological measurement of the object surface. For example, the surface roughness of features may be affected and changed by inaccurate edge preservation. 

Fig.~\ref{fig:laplacian-sketch} shows a sketch of how the MSE and the MSLE calculate misalignment in the features of an image. For this example, the image feature is represented as a square that is slightly shifted in position in image B from its original position in image A. The MSE calculates the difference between the two images and does not count the section of misalignment that is overlapping in the two images, and thus calculates a smaller value for the overall error as compared to the MSLE which takes into account the edges of both original image and decoded image. Thus, the MSLE is potentially more sensitive to edge feature changes. By comparing the MSLE for different image configurations, we can identify when significant changes to fine features occur. The calculated values for the MSLE of binary decoded images are shown in Table~\ref{table:binary-quality} where the loss in feature preservation between the large and small compression rates is emphasized.

Figure~\ref{fig:laplacian} shows the calculated Laplacian for the original image and decoded images obtained for the models and compression rates studied. Also shown is the difference image between the Laplacian of the decoded images and the original image. The difference images can clearly show the increase in feature preservation error as the compression rate also increases. The two models perform similarly for the smaller compression to $128 \times 128 \times 8$, while at the larger compression to $64 \times 64 \times 4$ the VQ-VAE model preserves edges and features with greater precision. As a result, if the goal is to preserve fine features for later analysis and recovery, then using smaller compression rates is recommended.

\section{Conclusions}

In this study, we applied deep learning autoencoder models to XCT images of sandstone samples to evaluate how such models preserve image features that are of importance to scientists.
The models used were D-CNN and VQ-VAE architectures. We also considered three different compression rates where the original images with size $512 \times 512$ were encoded to $128 \times 128 \times 8$, $64 \times 64 \times 4$, and $32 \times 32 \times 2$ arrays. The largest compression to a $32 \times 32 \times 2$ encoded array completely failed to decode the images in the D-CNN model and performed poorly for the VQ-VAE model resulting in blurry images.

Both models performed similarly for the small compression to $128 \times 128 \times 8$.
At the small compression rate, features were well preserved with low error in quality metrics and similar material properties, namely the total porosity percentage. However, as expected the VQ-VAE outperformed the D-CNN model at the compression to $64 \times 64 \times 4$ since the total number of trainable parameters increased significantly at this compression rate for the VQ-VAE model as compared to the D-CNN model.

The traditionally used metrics of quality assessment, namely the MSE and PSNR were found to not entirely capture how the decoded images differed from the original image. We proposed the use of a new metric called the MSLE that can evaluate the precision of how well the fine structure and edges of image features are preserved. The MSLE was a useful tool to clearly distinguish how features blur or change at larger compression rates when compared to smaller compression rates. Such evaluation and metrics can help scientists determine what model and compression rate to choose for preserving the features and calculations that are important for later analysis. For example, using smaller compression rates if detail preservation is important or using D-CNN for larger compression rates if detail preservation is not critical and global parameters such as total porosity are sufficient for analysis after decoding.

Future work can further explore the MSLE and metrics sensitive to edge detection by considering model performance when such a metric is included in the model loss function. Furthermore, by including additional training data from different samples of the same material, it would be interesting to investigate the creation of a general model for the compression of specific materials without the need to train on each individual sample individually. Such a general model in the case of a VQ-VAE architecture can also possibly help generate new synthetic images that can be used in instances when existing data is insufficient for further tasks such as segmentation or classification~\citep{Tsamos2019}.

\section*{Data availability}
The data and codes used in this work are available upon request.

\begin{ack}

We would like to sincerely thank the members of the X-ray Imaging division at BAM for their technical support and insightful discussions. We would also like to thank Sabine Kruschwitz for providing us with the Kylltaler sandstone sample. This work was supported by Deutsche Forschungsgemeinschaft grant no. 505623442.

\end{ack}

\bibliography{mybibfile}

\end{document}